\documentclass[journal=jacsat,manuscript=article]{achemso}
\setkeys{acs}{maxauthors = 180}

\usepackage[version=3]{mhchem} 
\usepackage{xcolor}

\author{Liam A. V. Nagle-Cocco}
\affiliation
{Stanford Synchrotron Radiation Lightsource, SLAC National Accelerator Laboratory, Menlo Park, California, 94025, United States of America}

\author{Andreas Schneemann~\footnote{Present address: Chair of Inorganic Chemistry I, Faculty of Chemistry and Food Chemistry, TU Dresden,01062 Dresden, Germany}~}
\affiliation
{Sandia National Laboratories California, Livermore, California, United States of America}

\author{Kevin H. Stone}
\affiliation
{Stanford Synchrotron Radiation Lightsource, SLAC National Accelerator Laboratory, Menlo Park, California, 94025, United States of America}

\author{Vitalie Stavila}
\affiliation
{Sandia National Laboratories California, Livermore, California, United States of America}

\author{Thomas Gennett}
\affiliation
{National Renewable Energy Laboratory, Golden, Colorado, United States of America}
\altaffiliation
{Chemistry Department, Colorado School of Mines, Golden, Colorado, United States of America}

\author{Nicholas A. Strange}
\affiliation
{Stanford Synchrotron Radiation Lightsource, SLAC National Accelerator Laboratory, Menlo Park, California, 94025, United States of America}
\email{nstrange@slac.stanford.edu}

\title
{Orientational Disorder of NH$_3$ in Hexammine Magnesium Borohydride}


\begin{document}

\begin{tocentry}

    \includegraphics[scale=1]{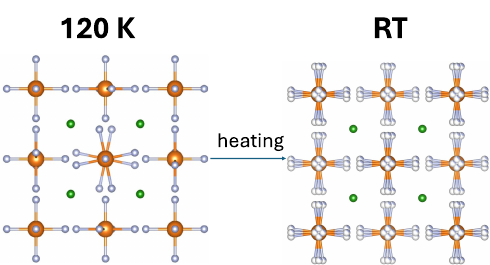}

For Table of Contents only.

\end{tocentry}

\begin{abstract}
  Hexammine magnesium borohydride\textcolor{black}{, Mg(NH$_3$)$_6$(BH$_4$)$_2$,} consists of adducted NH$_3$ molecules locked in a matrix of Mg cations and borohydride anions. It is a candidate material for hydrogen storage, with 16.8wt\% hydrogen stored in both the NH$_3$ and borohydride anions. It also may be of interest as a Mg$^{2+}$ conducting electrolyte in solid state batteries. Its crystal structure has, until now, eluded a proper structural solution due to ambiguity regarding the NH$_3$ position and behaviour. In this work, we show using synchrotron \textcolor{black}{X}-ray diffraction that the room-temperature structure can be solved only with a model assuming orientational disorder of ammonia molecules within the crystal structure. Cooling the sample to 120\,K yields additional Bragg peaks, which can only be solved with a unit cell expansion consistent with a freezing of the orientational freedom of ammonia molecules. Using this insight from the structure solution, we perform a full assignment of the vibrational modes in the room-temperature IR spectrum.  
\end{abstract}

\section{Introduction}

Complex metal hydrides are widely considered leading contenders for materials-based hydrogen storage~\cite{Klebanoff2012,schneemann2018nanostructured,kumar2019solid,chang2021emerging}. Second and third row complex metal hydrides offer high gravimetric and volumetric hydrogen storage capacities which partially meet targets established by the United States Department of Energy for materials-based hydrogen storage~\cite{Klebanoff2012,HFTO_MYPP_2024}. As promising as complex metal hydrides may appear, limiting factors such as slow dehydrogenation kinetics, high decomposition temperatures, and poor reversibility encumber their practical use as a hydrogen storage medium. As a result, several new approaches aimed at improving complex metal hydride material properties have been examined including the use of additives~\cite{newhouse2010reversibility,zhang2013thermal,zavorotynska2015hydrogen,leick2021al2o3,strange2022reactive}, composite materials~\cite{wang2017breaking}, nanoscaling/nanoconfinement~\cite{white2020melting,schneemann2020nanoconfinement,cho2021reversing}, and the search for new materials~\cite{allendorf2022challenges}. The mixture of hydridic and protic hydrogens into a single material is one route previously explored for improving the hydrogen cycling properties of hydrides. Typically, this involves the combination of a hydride and a material containing N-H (i.e., containing hydrogen with partial positive charge) bonds as a composite, adduct, molecular additive~\cite{kang2008ammonia,kumar2019solid}. 
The main goal is to redirect decomposition pathways so that it avoids forming stable byproducts, which can trap the system in low-energy states and make it difficult or impossible to reverse the reaction and store hydrogen again. 
To date these modified materials have only exhibited limited success for applications, since ammonia is a common byproduct of the decomposition reaction which is detrimental to hydrogen fuel cell using existing technology~\cite{yuan2012diagnosis,misz2016sensitivity}.

Conversely, materials-based ammonia storage is gaining momentum, due to developing processes and technologies which aim to use ammonia as a direct fuel source or a hydrogen carrier. As a fuel source, ammonia can run in fuel cells analogous to hydrogen~\cite{afif2016ammonia,jeerh2021recent}. As a hydrogen carrier, ammonia can be decomposed via catalysis processes which are the focus of much active research~\cite{lucentini2021review,weidenthaler2022situ,ulucan2024unveiling}. Although domestic transport of ammonia is available through existing pipelines in the U.S. and abroad~\cite{kass2023assessing}, the toxic nature of the gas adds complexity to its deployment as a widespread energy source. As regulations are being developed around the ammonia supply chain, efforts are underway to advance materials-based (e.g., solid state) storage to mitigate the risks and hazards associated with ammonia~\cite{chang2021emerging,kojima2020ammonia,guo2023efficient}.

Recent literature has demonstrated that a variety of mixed borohydride-NH species are possible which include both ammonium and amine/amide derivatives. Synthesis typically involves solvated magnesium borohydride treated with gaseous ammonia at sub-zero temperatures. Roedern and Jensen (2015)~\cite{roedern2015ammine} demonstrated that liquid ammonia could be used to generate the hexammine complex for a selection of metal (M = Cr, Co, Fe) borohydrides. The hexammine magnesium borohydride adduct was identified almost 70 years ago~\cite{Kodama1958,Huff1958}, and a crystal structure solution was offered by Semenenko \textcolor{black}{\textit{et al}} (1975)~\cite{semenenko1975synthesis}. This material was later used a precursor in a recent publication by Soloveichik \textcolor{black}{\textit{et al}} (2008)~\cite{soloveichik2008ammine} for synthesizing a diammine variant, however the measured diffraction data could not be matched to the structure reported by Semenenko \textcolor{black}{\textit{et al}}~\cite{semenenko1975synthesis}; only the unit cell parameter was reported for this data.

Herein, two possible synthesis pathways and the complete room-temperature structure solution of Mg(NH$_3$)$_6$(BH$_4$)$_2$ are reported. Based on Rietveld refinements of synchrotron \textcolor{black}{X-ray Diffraction (XRD)} data, we demonstrate that the most satisfactory structure solution is obtained when the nitrogen is translationally displaced, causing positional and rotational disorder on NH$_3$ within the crystal structure. We further report XRD data at 120\,K in which additional, previously unobserved, diffraction peaks emerge, and show these are consistent with a “freezing in” of this NH$_3$ positional and orientational disorder. Finally, the vibrational dynamics of Mg(NH$_3$)$_6$(BH$_4$)$_2$ at room temperature were examined using infrared (IR) spectroscopy and gaps in prior mode assignments are made based on comparison to hexammine metal halide analogues. 

\section{Experimental}

\subsection{Synthesis of Mg(BH$_4$)$_2$ precursor}

First, Mg(BH$_4$)$_2$ precursor was obtained using a modified version of a previously reported procedure for synthesizing $\gamma$-Mg(BH$_4$)$_2$~\cite{zanella2007facile}. 
A dry Schlenk flask was equipped with a stirring bar, heated \textcolor{black}{\textit{in vacuo}}, flushed several times with argon as a protecting gas and filled with dry toluene (150\,mL) obtained from a solvent purification system (Innovation Technology PureSolv) and the flask was sealed with a septum. Neat BH$_3$·SMe$_2$ (175\,mmol, 17.5\,mL) was added via syringe, and the mixture was stirred for 10\,min. 
100\,mL of 1\,M Mg(Bu)$_2$ in heptane (100\,mmol) were added dropwise over 30\,min, upon which the reaction mixture turned cloudy. The reaction mixture is then stirred for two hours. In contrast to the synthetic procedures towards $\gamma$-Mg(BH$_4$)$_2$ reported by Schneemann \textit{et al} (2020)~\cite{schneemann2020nanoconfinement} the cloudy mixture did not settle, so that the solvent was removed from the mixture by condensing it into a cooling trap and afterwards heating it to 85 °C \textcolor{black}{\textit{in vacuo}} overnight. 
The obtained solid was considerably more fluffy in nature than $\gamma$-Mg(BH$_4$)$_2$ and featured spectral traces of alkyl groups (see supporting information), leading to the assumption that the intermediate is a form of porous Mg(BH$_4$)$_2$ featuring heptane residues within its pore space. It will hereafter be referred to as hept-Mg(BH$_4$)$_2$ to distinguish it from $\gamma$-Mg(BH$_4$)$_2$. The material was stored inside an Ar-filled glovebox until further manipulation.

\textcolor{black}{
\textbf{Hazard statement.} Mg(Bu)$_2$ is pyrophoric and the commercial Mg(Bu)$_2$ solutions in heptane can spontaneously ignite when exposed to moist from air or water. It needs to be handled under an argon or nitrogen protective environment such as a Schlenk line, and must be quenched before disposal. 
Ammonia is toxic and corrosive and liquid ammonia evaporates at -33.3$^\circ$C. For the experiments a dedicated Schlenk line for experiments with ammonia located in a well-ventilated fume-hood which was equipped with teflon fittings. All parts were carefully dried before the experiments. The reaction vessel is cooled with a mixture of dry ice and isopropanol. To prevent the build-up of pressure during heating to room temperature, the reaction vessel is equipped with a pressure release valve. After the reaction, the liquid ammonia is slowly warmed and carefully evaporated to the ventilation system.
}

\subsection{Synthesis of Mg(NH$_3$)$_6$(BH$_4$)$_2$}

500 mg of the prior reaction product hept-Mg(BH$_4$)$_2$ were placed inside a 250 mL Schlenk flask equipped with a stirring bar inside a glovebox and sealed with a septum. Outside the glovebox, the flask was connected to a Schlenk line and via a needle through the septum to an ammonia bottle. The flask with the precursor was then evacuated and cooled in a dry ice/isopropanol bath to -78 °C. Slowly NH$_3$ was condensed into the flask under stirring until around 50\,mL were inside the flask. Afterwards the mixture was kept at -78 °C for 3 hours and subsequently the NH$_3$ was slowly evaporated and the material dried \textcolor{black}{\textit{in vacuo}} at room-temperature overnight (an aliquot was also dried at elevated temperatures, but as shown in Figure~\ref{Figure_S1} this led to reduced phase purity). The materials were stored under argon until further manipulation. 

A second route for the synthesis was developed in which the ammonia was not condensed in, but the sample was just treated with gaseous ammonia at room temperature. For this, 200 mg of the prior reaction product were filled into a Schlenk tube equipped with a stirring bar inside a glovebox and sealed with a septum. The Schlenk tube was connected outside the glovebox via a Schlenk line with an ammonia bottle by sticking a long needle that could reach the bottom of the Schlenk tube. Via several flushing and evacuation cycles the Argon in the Schlenk tube was replaced by ammonia. After exposure of the material to roughly 5 bars of ammonia for 3 hours, the Schlenk tube was evacuated overnight and the material was stored inside the glovebox for further manipulation. The reaction product was also obtained quantitatively and handled without further purification.

\subsection{X-ray diffraction}

In situ synchrotron radiation powder \textcolor{black}{X}-ray diffraction measurements were performed at \textcolor{black}{the Stanford Synchrotron Radiation Lightsource (SSRL)} beam line 2-1~\cite{stone2023remote} with $\lambda =0.729$\,\AA{}. Samples were loaded into 1\,mm diameter high-purity quartz capillaries with 10\,$\mu$m wall thickness (Charles Supper). 
The capillaries were sealed into a heating cell described by Hoffman \textcolor{black}{\textit{et al}} (2018)~\cite{hoffman2018situ}. The sample cell was then connected to a custom gas-handling manifold assembled at the \textcolor{black}{National Renewable Energy Laboratory}. For \textcolor{black}{\textit{in situ}} heating measurements, the sample cell was evacuated and subsequently backfilled with 5 bar H$_2$. Low-temperature measurements were conducted in ambient pressure argon and utilized an Oxford Cryostream to cool the samples to $\sim$120\,K using a narrow stream of liquid nitrogen. 2D diffraction images were captured on a portrait-oriented Pilatus 100K hybrid photon counting detector and subsequently integrated using a Python script specifically developed for the angle-dispersive two-circle configuration at SSRL beam line 2-1.

\subsection{X-ray analysis}

Rietveld~\cite{rietveld1969profile,rietveld2014rietveld} and Pawley refinements~\cite{pawley1981unit} were performed using TOPAS Academic, version 7~\cite{coelho2018topas}. For the XRD data at 120\,K, structure solution was performed following the route of Coelho (2017)~\cite{coelho2017indexing} assuming a cubic or rhombohedral cell. Rietveld refinements use the Thompson-Cox-Hastings pseudo-Voigt peakshape~\cite{thompson1987rietveld}.

\subsection{FTIR spectroscopy}

Fourier transform infrared spectroscopy (FTIR) measurements were conducted using attenuated-total-reflectance (ATR) sampling on an Agilent Cary 630 maintained in an argon glovebox.

\subsection{NMR spectroscopy}

Solid-state NMR spectroscopy experiments were performed at 11.7\,T (500.18\,MHz for $^1$H and 160.48\,MHz for $^{11}$B) on an Agilent VNMRS spectrometer at the Environmental Molecular Sciences Laboratory (EMSL), Pacific Northwest National Laboratory (PNNL), using a home-built 5\,mm MAS probe double tuned to $^1$H/$^{11}$B. 
The rotors were CAVERN-style zirconia sleeves (revolution NMR that were modified to accommodate a double O-ring and a Vespel bushing to seal the sample up to 225\,bar at 250\,$^\circ$C) that have been described previously by Walter \textcolor{black}{\textit{et al}} (2018)~\cite{walter2018operando}. 
Chemical shifts were referenced to 1 M boric acid (aq.) at 19.2\,ppm, and nonselective $\pi$/2 pulses were calibrated as 5.2\,$\mu$s on the same sample. 
Borohydride samples were packed into the rotors in a glovebox under an inert nitrogen atmosphere and sealed with the O-ring seals mentioned above. 
For quantitative direct polarization (DP) experiments, the radio frequency pulse was reduced to a selective $\pi$/20 or 0.26\,$\mu$s. The recycle delays were 1\,s. The $^1$H decoupling scheme was SPINAL-1657 with a field strength of approximately 42\,kHz.

\section{Results}

\section{Characterization of Mg(NH$_3$)$_6$(BH$_4$)$_2$ reaction pathways}

\begin{figure}[h]
    \includegraphics[scale=0.4]{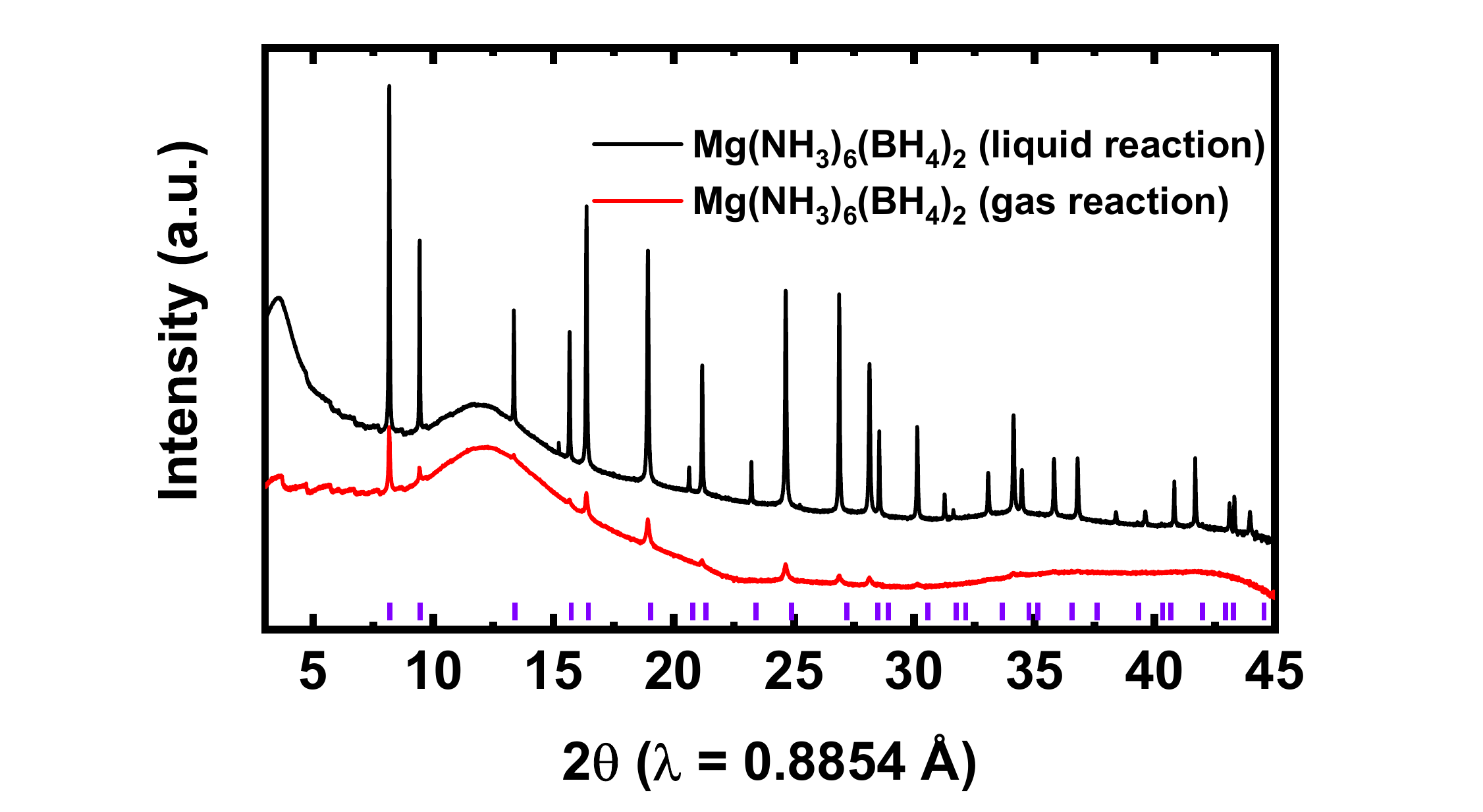}
    \caption{
        \label{Figure_1}
        Synchrotron XRD (from \textcolor{black}{beam line 2-1, Stanford Synchrotron Radiation Lightsource}) of Mg(NH$_3$)$_6$(BH$_4$)$_2$ synthesized by reaction between the hept-Mg(BH$_4$)$_2$ precursor and gaseous (red) or liquid (black) ammonia. 
        \textcolor{black}{Tick-marks correspond to $Q$-values where the Bragg condition is met for the reported $Fm\bar{3}m$ room-temperature structure.} 
    }
\end{figure}

After the precursor hept-Mg(BH$_4$)$_2$ material was reacted with ammonia, the hexammine complex formed. Sample treatment with either liquid or gaseous ammonia generates the adduct as deduced by the diffraction patterns shown in Figure~\ref{Figure_1}. However, the former method appears to produce a material with higher total conversion to the hexammine phase. Diffraction from the liquid ammonia-treated sample also includes a strong amorphous peak just below $2\theta=5$°. Neither material displays residual reflections from the Mg(BH$_4$)$_2$ precursor. Cross-polarization $^{11}$B NMR, displayed in Figure~\ref{Figure_S2}a, demonstrates that the broad distributions of boron environments previously observed in the intermediate material are largely reduced (with the exception of a subtle signal from borates) and only the borohydride environment is observed with appreciable intensity. As a result, introduction of ammonia appears to facilitate the conversion from a partially alkylated borohydride to the tetrahydroborate (BH$_4$)$^-$ anion. The $^{13}$C NMR spectrum of Mg(NH$_3$)$_6$(BH$_4$)$_2$, observed in Figure~\ref{Figure_S2}b, suggests that some amount of Mg(BH$_4$)$_2$ precursor or a derivative is still present, though the material remains amorphous and is accounted for by the broad amorphous signal in the synchrotron XRD [Figure~\ref{Figure_1}].

\section{Room temperature structure of Mg(NH$_3$)$_6$(BH$_4$)$_2$}

\begin{figure}[h]
    \includegraphics[scale=1]{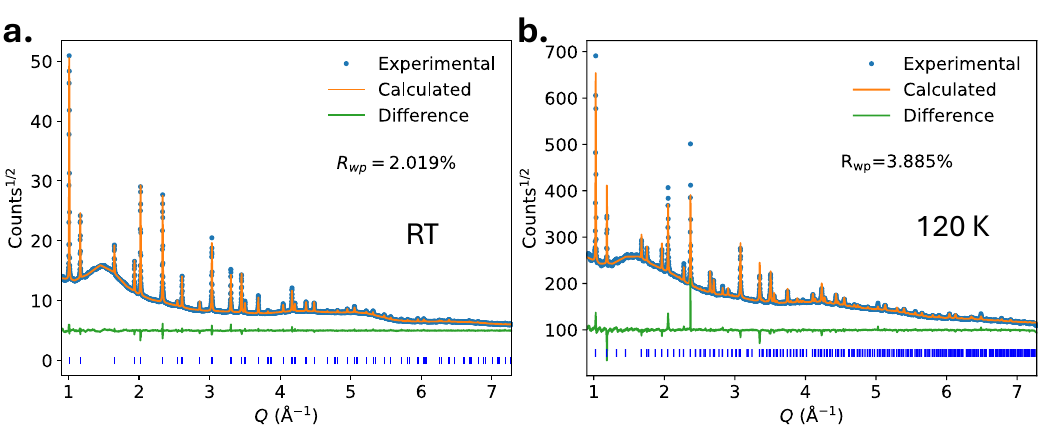}
    \caption{
        \label{Figure_2}
        Rietveld fit of XRD data from Mg(NH$_3$)$_6$(BH$_4$)$_2$ at (a) room-temperature using a cubic $Fm\bar{3}m$ cell with orientational and positional NH$_3$ disorder, and (b) 120\,K using a $2\times2\times2$ expansion of the room-temperature cubic cell with NH$_3$ ordering and \textcolor{black}{$Fm\bar{3}c$} space group. 
        In both figures, vertical tick-marks indicate allowed Bragg reflections for the structure.
    }
\end{figure}

\begin{figure}[h]
    \includegraphics[scale=0.95]{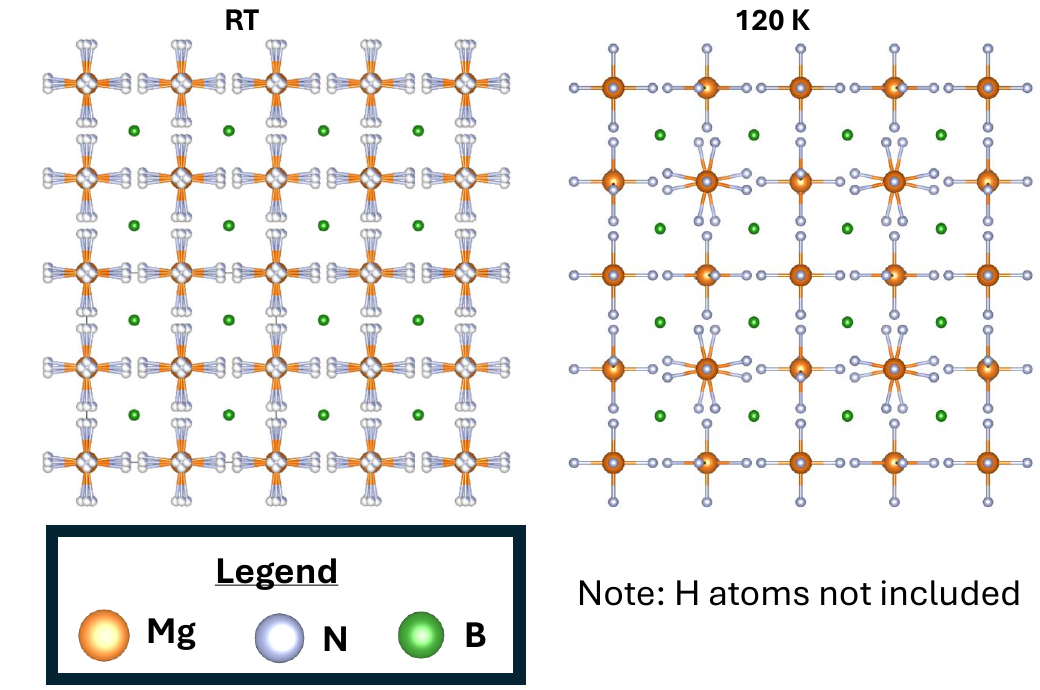}
    \caption{
        \label{Figure_3}
        Side view of the Mg(NH$_3$)$_6$(BH$_4$)$_2$ unit cell at room-temperature (left), with orientational and positional disorder of N sites (representing ammonia molecules) and 120\,K (right) where the N sites are ordered, manifesting as MgN$_6$ octahedral tilts. 
        Hydrogen atoms are excluded to reflect precisely the output of Rietveld refinement. 
        The room-temperature unit cell is doubled in each spatial direction for direct comparison with the 120\,K unit cell. 
        B: green; N: white; Mg; orange.
    }
\end{figure}

The structure of the liquid ammonia-treated sample was studied using synchrotron \textcolor{black}{X}-ray diffraction. 
Since the room-temperature structure was previously indexed~\cite{soloveichik2008ammine} to a unit cell with $Fm\bar{3}m$ space group, Mg$^{2+}$ and BH$_4^-$ fractional positions were generated from an analogous Mn(BH$_4$)$_2$(NH$_3$)$_6$ structural model (ICSD 429495~\cite{jepsen2015tailoring}) and fixed. Given the similarity to hexammine metal halide structures, the initial coordinate for nitrogen was derived from the reported MgCl$_2$(NH$_3$)$_6$ structure (ICSD 290894~\cite{jones2013structure}). 
The converged fit to the room temperature experimental diffraction pattern is displayed in Figure~\ref{Figure_2}a with refinement results presented in Table~\ref{table1}.

The resulting magnesium borohydride hexammine complex exhibits a room temperature “disordered” K$_2$PtCl$_6$-type structure, visualized in Figure~\ref{Figure_3}, similar to the halide structural analogs such as Mg(NH$_3$)$_6$Cl$_2$~\cite{sorby2011crystal}. 
Magnesium atoms within the cubic lattice are “shielded” by ammonia molecules situated at the vertices of Mg(NH$_3$)$_6$ octahedra. 
Correspondingly, BH$_4$ anions are coordinated to the central magnesium cation at the magnesium ammine octahedral faces with BH$_4$ vertices oriented toward the central magnesium atom, leading to an apparent B-H-Mg-H-B linear structure. To simplify the structural model and compensate for poor sensitivity of X-rays to hydrogen atoms, the BH$_4$ positions and orientations were replicated from Mn(NH$_3$)$_6$(BH$_4$)$_2$ as reported by Jepsen \textcolor{black}{\textit{et al}} (2015)~\cite{jepsen2015tailoring}. 
\textcolor{black}{This results in H atoms located along the four B$\rightarrow$Mg directions.} 
Interestingly, preliminary simulated annealing cycles with subsequent refinements of the position and orientation of ammonia molecules suggested structures with hydrogen atoms projecting 4-fold symmetry about the nitrogen rather than the trigonal pyramidal molecular geometry of an ammonia molecule. 
However, it turns out that even in the Mg(NH$_3$)$_6$X$_2$ (X = halide) structures, two of three ammonia hydrogens orient along the edge of a projected square~\cite{jones2013structure}. 
As a result, the nitrogen atoms are pulled away from the 24$e$ Wyckoff site in the ideal structure and become fractionally occupied in the time-averaged structure. 
\textcolor{black}{This fractional occupancy indicates that there is a discrepancy between local and average structure, and this likely leads to disorder in BH$_4$ orientations in the average local structure, due to the electrostatic interaction between NH$_3$ and BH$_4$ groups.} 
As will be discussed later, the room-temperature hexammine structure exhibits considerable rotational and vibrational dynamics which leads to disorder in the K$_2$PtCl$_6$-type lattice.

The disordered structure of Mg(NH$_3$)$_6$(BH$_4$)$_2$ is unique relative to the other hexammine metal (M=Co, Fe, Mn) borohydride structures reported by Roedern and Jensen (2015)~\cite{roedern2015ammine}. Here, rather than ordering on the 24$e$ Wyckoff site with $4mm$ symmetry, the central nitrogen atom is translated slightly in the y fractional coordinate, which reduces the symmetry to $m$, on Wyckoff site 96$j$. This occurs as two of ammonia’s nitrogen atoms project towards adjacent BH$_4^-$ units resulting in a four-fold projected symmetry around the nitrogen. Each projection is only occupied one quarter of the time, resulting in positional and orientational disorder. The fit from this structural model is compared to the fit obtained with N on the higher-symmetry 24$e$ Wyckoff site in Figure~\ref{Figure_S3}; a far superior $R_\mathrm{wp}$ is obtained in the former case (2.021\% compared with 4.563\%).

\begin{figure}[h]
    \includegraphics[scale=0.6]{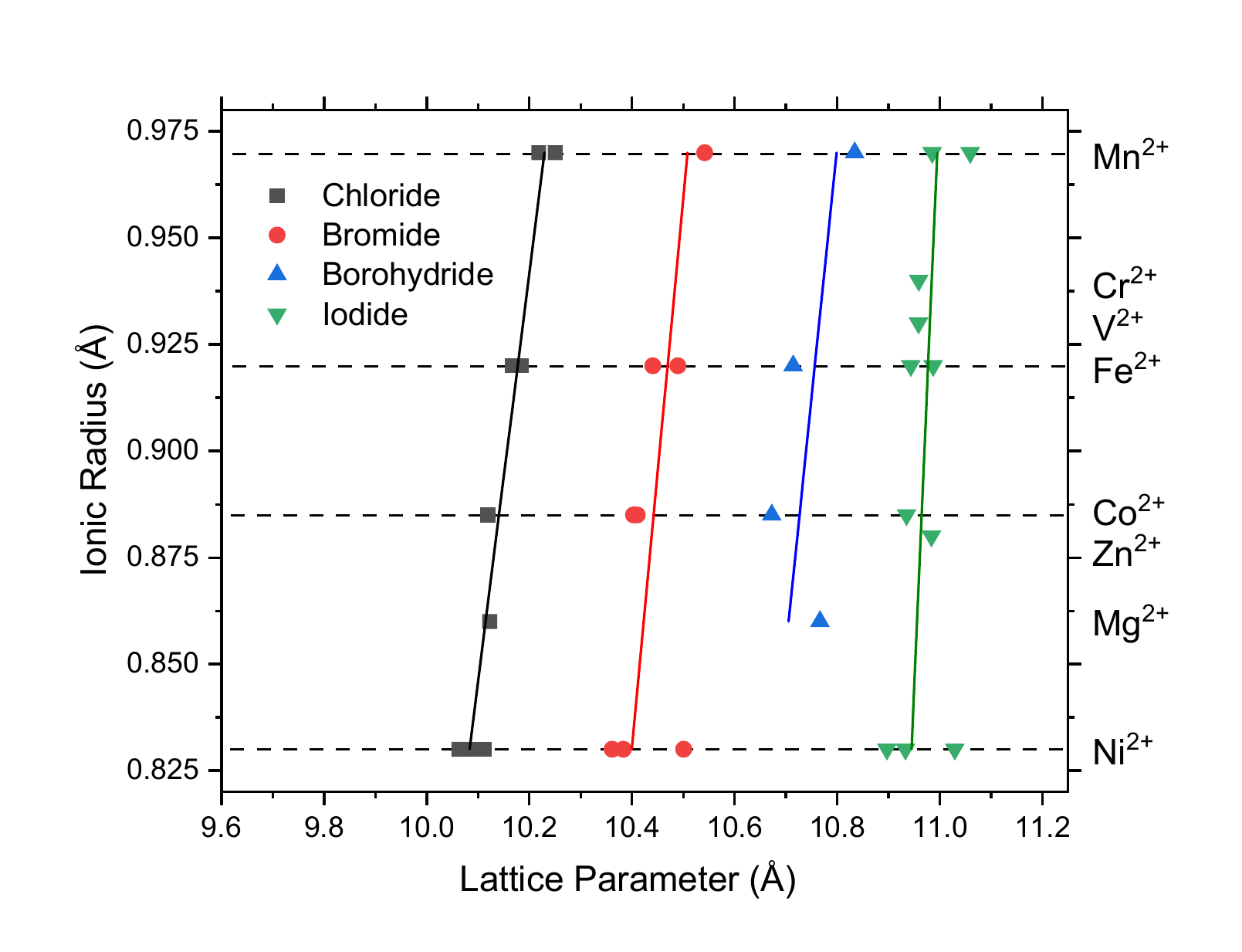}
    \caption{
        \label{Figure_4}
        Plot of ionic radius of M cation versus room-temperature lattice parameter of cubic unit cell for the series [M(NH$_3$)$_6$]X$_2$, where X is a halide or borohydride anion. 
        The [Mg(NH$_3$)$_6$](BH$_4$)$_2$ lattice parameter is as reported in this work, and the other borohydrides are taken from Ref.~\cite{roedern2015ammine}. 
        The remaining lattice parameters are taken from Eßmann \textcolor{black}{\textit{et al}} (1996)~\cite{essmann1996isotype}. Cationic radii are taken from Shannon (1976)~\cite{shannon1976revised}. 
        Lines of best fit are prepared via least squares fit in the x-axis.
    }
\end{figure}

Using the framework of Eßmann \textcolor{black}{\textit{et al}} (1996)~\cite{essmann1996isotype}, we can contextualize our derived Mg(NH$_3$)$_6$(BH$_4$)$_2$ structure among known M$^{2+}$(NH$_3$)$_6$X$_2$ (X = halides, borohydride) structures. Figure~\ref{Figure_4} has been reproduced from Eßmann \textcolor{black}{\textit{et al}} (1996)~\cite{essmann1996isotype}, and we have added data points for the Mn, Fe, and Co borohydride derivatives. 
The ionic radius reported for BH$_4^-$ is 2.05\,\AA{}~\cite{hino2012halide}. In comparison~\cite{shannon1976revised} to Br$^-$ (r = 1.96\,\AA{}) and I$^-$ (r = 2.06\,\AA{}), and only considering the effect of anionic radii, the M(NH$_3$)$_6$(BH$_4$)$_2$ structures should reside on an approximated line slightly lower in lattice parameter relative to the iodide derivatives. Although all the reported borohydride derivative lattice parameters are between bromide and iodide, there is some variation with the hexamine magnesium borohydride structure residing closer to the iodide than the other borohydride derivates. It is unclear how much of this is due to error in the lattice parameter determination or effects such as the presence/absence of the orientational disorder. 
This plot suggests there is potentially a series of cubic hexamine metal borohydrides that are yet to be synthesized (i.e. M=Cr, V, Zn, Ni), though there are almost certainly limits of cation size that bound the range of stability for this particular structure. Additionally, given that the estimated ionic radii for other complex hydride anions, such as AlH$_4^-$ and NH$_2^-$,~\cite{simoes2017estimation} are significantly different than the halides, it is likely that borohydride is the only complex hydride that forms the K$_2$PtCl$_6$-type structure.

\begin{table}[h!]
\centering
\caption{
    Crystal structure data for the structure solutions at room-temperature and 120\,K of Mg(NH\textsubscript{3})\textsubscript{6}(BH\textsubscript{4})\textsubscript{2}. Atomic positions are given as unitless fractional coordinates. H positions are excluded given the unfeasibility of refining their positions against \textcolor{black}{X}-ray radiation. For the 120\,K data, there are two different Mg and B sites and three N sites. For the room-temperature data, there is just one site for each element.
    \label{table1}
}
\begin{tabular}{|l|c|c|cc|c|c|cc|}
\hline
\multicolumn{2}{|l|}{\textbf{Parameter}} & \textbf{Room Temp} & \multicolumn{6}{c|}{\textbf{120\,K}} \\
\hline
\multicolumn{2}{|l|}{Crystal system} & Cubic & \multicolumn{6}{c|}{Cubic} \\
\multicolumn{2}{|l|}{Space group} & $Fm\overline{3}m$ (225) & \multicolumn{6}{c|}{\textcolor{black}{$Fm\overline{3}c$ (226)}} \\
\multicolumn{2}{|l|}{$a$ (\AA)} & 10.7649(7) & \multicolumn{6}{c|}{\textcolor{black}{21.235(9)}} \\
\multicolumn{2}{|l|}{$\alpha$, $\beta$, $\gamma$ (°)} & 90, 90, 90 & \multicolumn{6}{c|}{90, 90, 90} \\
\multicolumn{2}{|l|}{$V$ (\AA$^3$)} & 1247.5(3) & \multicolumn{6}{c|}{\textcolor{black}{9575(12)}} \\
\multicolumn{2}{|l|}{$V$ per Mg atom (\AA$^3$)} & 311.875(6) & \multicolumn{6}{c|}{299.2(4)} \\
\hline
\textbf{Mg site} & Wyckoff & 4$a$ & \multicolumn{3}{c|}{24$d$} & \multicolumn{3}{c|}{\textcolor{black}{8b}} \\
 & (x,y,z) & (0,0,0) & \multicolumn{3}{c|}{(1/4, 1/4, 0)} & \multicolumn{3}{c|}{(0, 0, 0)} \\
 & Occupancy & 1 & \multicolumn{3}{c|}{1} & \multicolumn{3}{c|}{1} \\
\hline
\textbf{B site} & Wyckoff & 4$b$ & \multicolumn{6}{c|}{\textcolor{black}{64$g$}} \\
& (x,y,z) & (1/4,1/4,1/4) & \multicolumn{6}{c|}{(1/8, 1/8, 1/8)}  \\
& Occupancy & 1 &  \multicolumn{6}{c|}{1} \\
\hline
\textbf{N site} & Wyckoff & 96$j$ & \multicolumn{2}{c|}{\textcolor{black}{96$i$}} & \multicolumn{2}{c|}{\textcolor{black}{48$e$}} & \multicolumn{2}{c|}{\textcolor{black}{48$f$}} \\
& x & 0 &\multicolumn{2}{c|}{\textcolor{black}{0.7280(3)}} & \multicolumn{2}{c|}{0} & \multicolumn{2}{c|}{1/4} \\
 & y &  0.0289(5) & \multicolumn{2}{c|}{\textcolor{black}{0.6448(8)}} & \multicolumn{2}{c|}{0} & \multicolumn{2}{c|}{\textcolor{black}{0.6051(15)}} \\
 & z & 0.20077(15) & \multicolumn{2}{c|}{\textcolor{black}{1/2}} & \multicolumn{2}{c|}{\textcolor{black}{0.1049(14)}} & \multicolumn{2}{c|}{1/4} \\
& Occupancy & 1/4 & \multicolumn{2}{c|}{1} & \multicolumn{2}{c|}{1} & \multicolumn{2}{c|}{1} \\
\hline
\multicolumn{2}{|l|}{\textcolor{black}{$R_\mathrm{wp}$ (\%)}} & \textcolor{black}{2.019} & \multicolumn{6}{c|}{\textcolor{black}{3.885}} \\
\hline
\end{tabular}
\end{table}

\section{Loss of orientational and positional NH$_3$ disorder on cooling to 120\,K}

Upon cooling to 120\,K, several new peaks emerge in the diffraction pattern. 
As this does not occur in tandem with a change in the peak structure of the room-temperature diffraction pattern, we interpreted this in terms of a supercell expansion rather than desymmetrisation of the unit cell. 
Indexing the complete set of peaks assuming a cubic cell, using the approach of Coelho (2017)~\cite{coelho2017indexing}, suggested a $2\times2\times2$ expansion of the unit cell with space group symmetry \textcolor{black}{$Fm\bar{3}c$}.

Seeking to determine whether the supercell expansion was due to ordering of the NH$_3$ or BH$_4$ molecular units, we then performed simulated annealing on this cubic \textcolor{black}{$Fm\bar{3}c$} supercell against the data. In this simulated annealing, N and B atomic positions were subject to randomization, and ADPs were constrained such that B$_\mathrm{iso}$(B) $>$ B$_\mathrm{iso}$(N) $>$ B$_\mathrm{iso}$(Mg). 
Hydrogen was excluded due to the insensitivity of \textcolor{black}{X}-rays to hydrogen. The best resulting fit is shown in Figure~\ref{Figure_2}b and is solely due to ammonia molecules ordering in a $2\times2\times2$ cell expansion with a regular tilting of MgN$_6$ octahedra, with B atom positions refining to the high-symmetry inter-octahedral sites as in the room-temperature structure [Figure~\ref{Figure_3}\textcolor{black}{]}.

\textcolor{black}{
This N-ordering results in a range of B-N distances. For each B atom, there are four MgN$_6$-B interactions where three N atoms are near the B cation. One of these has three equal B-N distances with length $\sim$3.778\,\AA{}. The other three MgN$_6$-B interactions each have three unique B-N distances: $\sim$3.465\,\AA{}, $\sim$3.778\,\AA{}, $\sim$4.119\,\AA{}. These distances occur with $C_3$ rotational symmetry about the axis passing through the Mg-B distance with three equal B-N distances. 
While we cannot prove this without performing neutron diffraction on a deuterated sample (due to the negligible x-ray scattering of hydrogen), we can speculate that this N ordering will consequently lead to an ordered arrangement of BH$_4$ molecular units. This likely manifests as a rotation of BH$_4$ units about the Mg-B axis with three equal B-N distances.
}

We note that calculated peak intensities do not perfectly match the experimental intensities, particularly that for the [800] peak ($Q\approx2.37$\,\AA{}$^{-1}$) which is anomalously narrow. We cannot achieve a perfect fit even via the Pawley approach~\cite{pawley1981unit} with this and \textcolor{black}{$Fm\bar{3}c$} unit cell [Figure~\ref{Figure_S4}], due to the [800] peakshape, and the Stephens approach~\cite{stephens1999phenomenological} to modelling this also fails to achieve a perfect fit in conjunction with the cubic \textcolor{black}{$Fm\bar{3}c$} unit cell proposed here. 
\textcolor{black}{Removing the centre of symmetry does not resolve the issue, as seen by Rietveld refinement with the $F\bar{4}3c$ space group [Figure~\ref{Figure_S5.pdf}].} 
\textcolor{black}{We also tested a Rietveld refinement assuming both the room-temperature $Fm\bar{3}m$ phase and the low-temperature $Fm\bar{3}c$ phase [Figure~\ref{SI_dual-Rietveld-120K.pdf}].} 
\textcolor{black}{It is possible that there is some subtle desymmetrisation of the unit cell.} Such desymmetrisation is unlikely to be a tetragonal or orthorhombic distortion along the cubic unit cell directions, as peaks such as the [800] would be expected to broaden in such a case, rather than narrow.

A previous neutron diffraction study on deuterated Mg(ND$_3$)$_2$X$_2$ (X=halide) compositions~\cite{leineweber2001uniaxial}, which are not directly comparable to the material studied in this work due to the reduced number of ammonia molecules per formula unit, indicated that there is an order-disorder transition in which orientational disorder of the ammonia molecules at room-temperature (manifesting as four-fold rotational symmetry about the Mg-N bond axis) disappears on cooling to base temperatures. 
While we cannot directly test this hypothesis without deuterated samples and neutron diffraction, whether or not an analogous transition occurs in the presently-studied Mg(NH$_3$)$_2$(BH$_3$)$_2$ material depends on whether there is free, unhindered rotation of the NH$_3$ molecule about the Mg-N bond at room-temperature. 
As we will discuss later, we suspect this is not the case and so a directly analogous order-disorder transition likely does not occur. However, we do interpret our findings as indicating a similar, but distinct order-disorder transition in the positional disorder of NH$_3$ molecules.

\section{Effect of heating on the structure of Mg(NH$_3$)$_6$(BH$_4$)$_2$}

\begin{figure}[h]
    \includegraphics[scale=1]{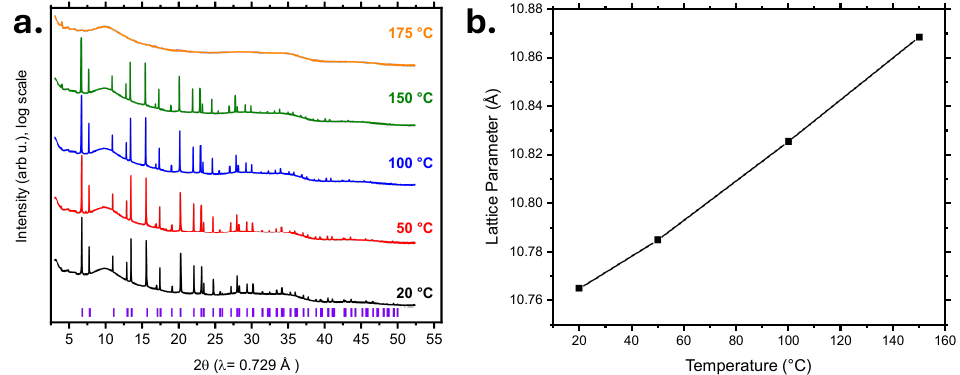}
    \caption{
        \label{Figure_5}
        a) Temperature dependent XRD of Mg(NH$_3$)$_6$(BH$_4$)$_2$ and b) the temperature dependence of the lattice parameter extracted from the XRD data by Rietveld refinement using the room-temperature $Fm\bar{3}m$ structure. Here, heating occurs under a 5 bar pressure of H$_2$. 
        \textcolor{black}{Tick-marks correspond to $Q$-values where the Bragg condition is met for the reported $Fm\bar{3}m$ room-temperature structure.} 
        In (b) the line connects the points as a guide to the reader.
    }
\end{figure}

\textit{In situ} XRD was performed on the Mg(NH$_3$)$_6$(BH$_4$)$_2$ material during heating under 5\,bar hydrogen backpressure in an attempt to prevent thermal decomposition, by analogy with the enhanced thermal stability of the base material $\gamma$-Mg(BH$_4$)$_2$ under hydrogen pressure reported by White \textcolor{black}{\textit{et al}} (2020)~\cite{white2020melting}. 
Previous studies~\cite{plevsek1966chemistry,konoplev1985ammonia,soloveichik2008ammine} reported that the Mg(NH$_3$)$_6$(BH$_4$)$_2$ structure decomposes entirely on heating between 150 and 175 °C, whereas heating to lower temperatures in dynamic vacuum yields the diammoniate complex. 
Under the 5 bar hydrogen backpressure, we see precisely the same decomposition reported by Soloveichik \textcolor{black}{\textit{et al}} (2008)~\cite{soloveichik2008ammine} for atmospheric heating, as shown in Figure~\ref{Figure_5}. These results suggest that hydrogen pressures greater than 5\,bar may be required to stabilize the material, which is not surprising given that White \textcolor{black}{\textit{et al}} (2020)~\cite{white2020melting} used hydrogen pressures of 100s of bar to stabilize against decomposition in the $\gamma$-Mg(BH$_4$)$_2$ system.

\section{Vibrational Dynamics of Mg(NH$_3$)$_6$(BH$_4$)$_2$}

\begin{figure}[t]
    \includegraphics[scale=0.48]{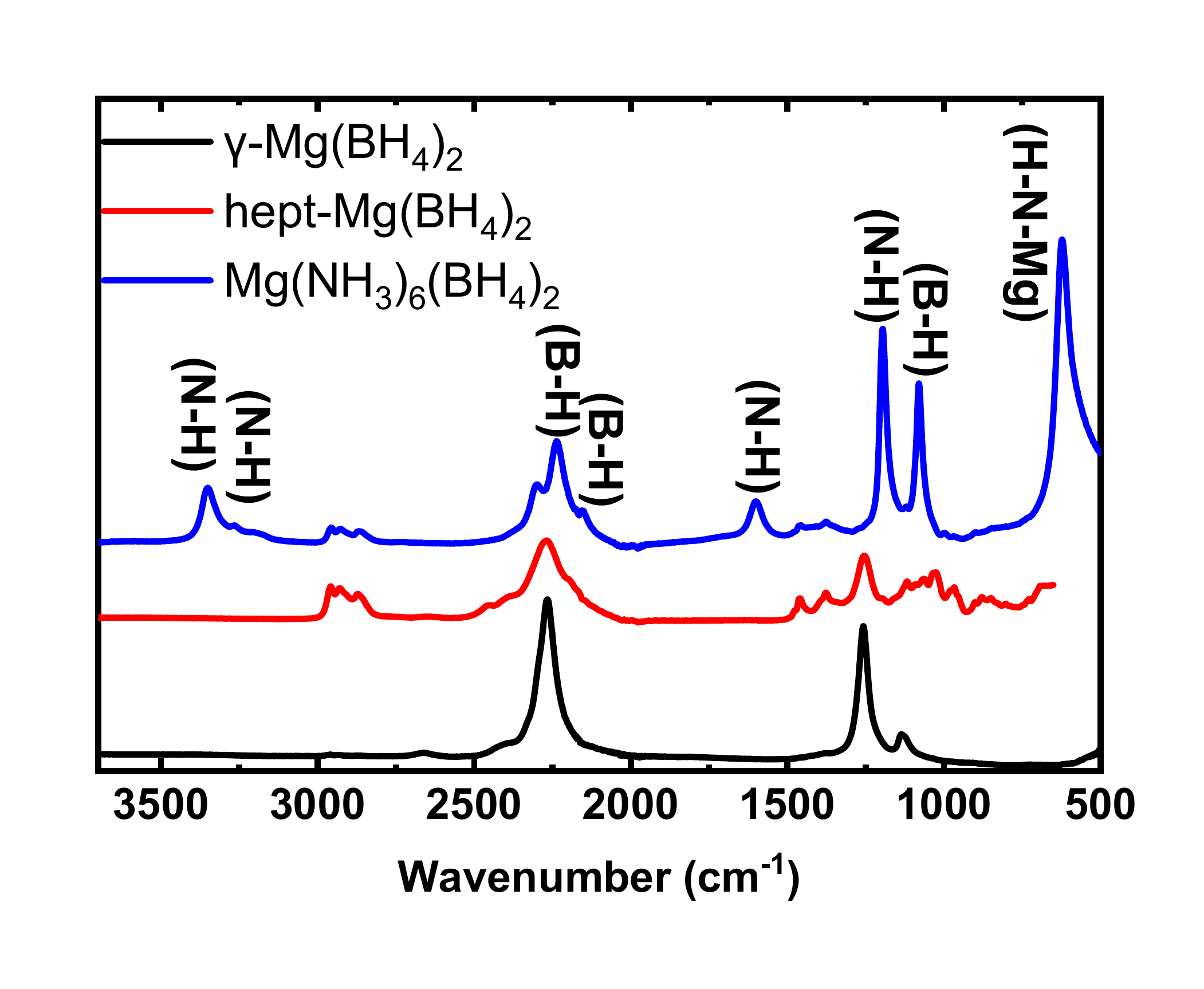}
    \caption{
        \label{Figure_6}
        FTIR of $\gamma$-Mg(BH$_4$)$_2$ (black) as reported previously~\cite{schneemann2020nanoconfinement}, hept-Mg(BH$_4$)$_2$ (red), and Mg(NH$_3$)$_6$(BH$_4$)$_2$ (blue). 
        \textcolor{black}{Rotated bold bond descriptions in brackets indicate the assignment of Mg(NH$_3$)$_6$(BH$_4$)$_2$ vibrations responsible for the labelled peak, as described in Table~\ref{table2}}
    }
\end{figure}

The dynamics of the hexammine complex, examined using FTIR, are displayed in Figure~\ref{Figure_6}. 
It is not too surprising that the FTIR spectrum exhibits a number of similarities to that observed from the halide structures, in particular for the dynamics of ammonium molecules. 
The IR vibrational modes from Mg(NH$_3$)$_6$(BH$_4$)$_2$ find strong agreement with N-H dynamics observed in other M(NH$_3$)$_6$X$_2$ (M = Co, Ni, Fe; X = Cl, Br, I) complexes~\cite{jacobs1987rontgenographische,grzybek1973spectroscopic,reardon2012ammonia,essmann1996isotype}. However, as expected, the substitution of a borohydride anion for the halide causes some shifts in excitation wavenumbers. 
In general, one might expect the NH$_3$ modes influenced by the coordinated anions as a hindering potential to shift to higher energies for smaller radii and higher electronegativity (e.g., F$^-$ $>$ Cl$^-$ $>$ Br$^-$ $>$ I$^-$). 
Ignoring electronegativity arguments, anion size has the most prominent effect on lattice parameter and as a result H-X distances. Peak assignments for the N-H modes are provided in Table~\ref{table2}. Frequencies not listed in the table are attributed to modes from residual the heptane-containing MgBH$_4$ precursor. The excitations between 2000 and 2500 cm$^{-1}$ are the result of B-H stretching modes and are split in the hexammine complex relative to the singlet observed in $\gamma$-Mg(BH$_4$)$_2$ (black trace) near 2266 cm$^{-1}$.

The dynamics of NH$_3$ in any M(NH$_3$)$_6$X$_2$ complex will be influenced by the associated interatomic potentials (M-N and N-H-X) and distances. Previous reports on M(NH$_3$)$_6$I$_2$ (M = Fe, Mn) have indicated that NH$_3$ rotations about the M-N axes are possible~\cite{reardon2012ammonia,tekin2010ammonia} and can be characterized as free rotors at room temperature~\cite{reardon2012ammonia,essmann1996isotype}. The latter assertion may be an overinterpretation of data provided by Jacobs \textcolor{black}{\textit{et al}}~\cite{jacobs1987rontgenographische}, where the authors indicate that stretching ($\nu_s$, $\nu_a$) and deformation ($\delta_s$, $\delta_d$) bands observed for M(NH$_3$)$_6$I$_2$ (M = Mn$^{2+}$, Fe$^{2+}$) in the IR spectrum are very broad, combined with an intense (H-N-M) librational mode ($\delta_e$). They associate these observations with rotational disorder of the NH$_3$ group where the protons are located at maxima of a probability distribution. 
They later contend that based on the H positions refined from single crystal XRD, the NH$_3$ rotations are unhindered. If NH$_3$ exhibited free rotations in this complex, an equal probability of H positions would be observed about the NH$_3$ C$_3$ axis in structural data. Although Jacobs obtained single-crystal XRD of the M(NH$_3$)$_6$I$_2$ compounds, the accuracy (and associated uncertainty) of the H positions is not made clear and would benefit from neutron diffraction, which is sensitive to the positions of deuterium. Schiebel \textcolor{black}{\textit{et al}} (1993)~\cite{schiebel1993orientational} presented neutron diffraction data indicating orientational disorder in the Ni(NH$_3$)$_6$I$_2$ system and proposed the results provided evidence of rotation-translation coupling. Their ``alternative” structural model (i.e., the one used in the current study for Mg(NH$_3$)$_6$(BH$_4$)$_2$) was later adapted for the Mg(NH$_3$)$_6$Cl$_2$ system and validated by comparison to powder neutron diffraction data and first principles molecular dynamics~\cite{sorby2011crystal}. It is peculiar that the NH$_3$ orientational disorder is present for the Ni(NH$_3$)$_6$I$_2$ system, but not reported for other metal ammonia complexes. It is possible that the small cation radius paired with a large anion is sufficient to induce disorder based on interatomic spacing considerations alone. Alternatively, perhaps this feature is a subtle detail overlooked in many of these structures (for instance, Figure~\ref{Figure_S3} shows that a reasonable, but inferior, fit to the room-temperature diffraction data can be achieved without orientational disorder).

For the Mg(NH$_3$)$_6$(BH$_4$)$_2$ system, it would be valuable to supplement the existing XRD data with neutron diffraction and quasi-elastic neutron scattering in order to better understand the relationship between structural disorder and dynamics (e.g., rotation-translation coupling dependence on temperature and anion, particularly when comparing the monodentate N-H-Cl and bidentate N-H-H$_2$-B interactions). Although the XRD results indicate the N-H-H-B spacing is proportional to the anion radius (i.e. in between Br$^-$ and I$^-$), BH$_4^-$ may exhibit a stronger potential with the NH$_3$ protons (relative to other metal-halide pairs) due to bidentate proton-hydride interactions, which potentially induce the positional and orientational disorder.

\begin{table}[htbp]
\centering
\caption{
    \label{table2}
    Assignment of FTIR vibrational frequencies from Mg(NH$_3$)$_6$(BH$_4$)$_2$. ($\nu_s$, $\nu_a$ = symmetric, asymmetric stretch; $\delta_s$, $\delta_d$ = symmetric, degenerate bend). References used for the assignment of Mg(NH$_3$)$_6$(BH$_4$)$_2$ modes are provided in the Mg(NH$_3$)$_6$(BH$_4$)$_2$ column.
}
\begin{tabular}{|l|c|c|c|}
\hline
\textbf{Mode} & \textbf{Mg(NH$_3$)$_6$(BH$_4$)$_2$} & \textbf{Mg(NH$_3$)$_6$Cl$_2$}  & \textbf{Mn(NH$_3$)$_6$(BH$_4$)$_2$} \\
 & \textit{(This work.)} & \textit{(Ref. \cite{plus1973etude})} & \textit{(Ref. \cite{jepsen2015tailoring})} \\
 & \textcolor{black}{(cm$^{-1}$)} & \textcolor{black}{(cm$^{-1}$)} & \textcolor{black}{(cm$^{-1}$)} \\
\hline
$\delta_d$, (H-N-Mg) & 623 \cite{jacobs1987rontgenographische} & 650 & 622 \\
$\delta$ (B-H) & 1080 \cite{soloveichik2008ammine} & – & 1081 \\
$\delta_s$ (N-H) & 1196 \cite{jacobs1987rontgenographische} & 1170 & 1193 \\
$\delta_d$ (N-H) & 1601 \cite{jacobs1987rontgenographische} & 1603 & 1595 \\
$\nu$ (B-H) & 2153 \cite{soloveichik2008ammine} & – & Not observed \\
$\nu$ (B-H) & 2236 \cite{soloveichik2008ammine} & – & 2241 \\
$\nu$ (B-H) & 2300 \cite{soloveichik2008ammine} & – & 2305 \\
$\nu_s$ (N-H) & 3265 \cite{jacobs1987rontgenographische} & 3210 & 3264 \\
$\nu_a$ (N-H) & 3350 \cite{jacobs1987rontgenographische} & 3353 & 3352 \\
\hline
\end{tabular}
\end{table}

\section{Conclusion}

In this work, we provide a comprehensive structural solution for Mg(NH$_3$)$_6$(BH$_4$)$_2$ which is consistent with the experimental diffraction pattern at room-temperature, showing that there is positional and orientational disorder of the ammonia molecules in the crystal structure. We further show that, on cooling, this positional disorder disappears by 120\,K, with the ammonia molecules freezing into an ordered arrangement of Mg(NH$_3$)$_6$ octahedral tilts analogous to that which occurs in perovskites, with a doubling of the unit cell in each direction. Additionally, we assign the full IR spectrum of this material. These findings will prove useful in further studies examining Mg(NH$_3$)$_6$(BH$_4$)$_2$ as a hydrogen storage material. 

\begin{acknowledgement}

This work was supported by the U.S. Department of Energy under Contract No. DE-AC36-08GO28308 with the Alliance for Sustainable Energy, LLC, the manager and operator of the National Renewable Energy Laboratory (NREL). 
Sandia National Laboratories is a multi-mission laboratory managed and operated by National Technology and Engineering Solutions of Sandia, LLC., a wholly owned subsidiary of Honeywell International, Inc., for the U.S. Department of Energy’s National Nuclear Security Administration (NNSA) under contract DE-NA-0003525. 
The use of the Stanford Synchrotron Radiation Lightsource, SLAC National Accelerator Laboratory, is supported by the US Department of Energy, Office of Science, Office of Basic Energy Sciences under Contract No. DE-AC02-76SF00515. 
The authors gratefully acknowledge engineering support from J.R. Troxel in configuring SSRL beam line 2-1~\cite{stone2023remote} for operation. 
The views and opinions of the authors expressed herein do not necessarily state or reflect those of the United States Government or any agency thereof. Neither the United States Government nor any agency thereof, nor any of their employees, makes any warranty, expressed or implied, or assumes any legal liability or responsibility for the accuracy, completeness, or usefulness of any information, apparatus, product, or process disclosed, or represents that its use would not infringe privately owned rights. 
Crystal structure figures were prepared using CrystalMaker$^\mathrm{TM}$~\cite{palmer2015visualization} and Vesta-III~\cite{momma2011vesta}.

\end{acknowledgement}

\begin{suppinfo}

SI contains additional FTIR and NMR data, and some additional Rietveld and Pawley refinements supporting the claims within the manuscript. (PDF)

\end{suppinfo}

\bibliography{achemso-demo}

\clearpage
\section{Supplementary Information}
        \renewcommand{\theequation}{S\arabic{equation}}
	\renewcommand{\thetable}{S\arabic{table}}
	\renewcommand{\thefigure}{S\arabic{figure}}
	\setcounter{equation}{0}
	\setcounter{figure}{0}
	\setcounter{table}{0}

\begin{figure}[h]
    \includegraphics[scale=0.4]{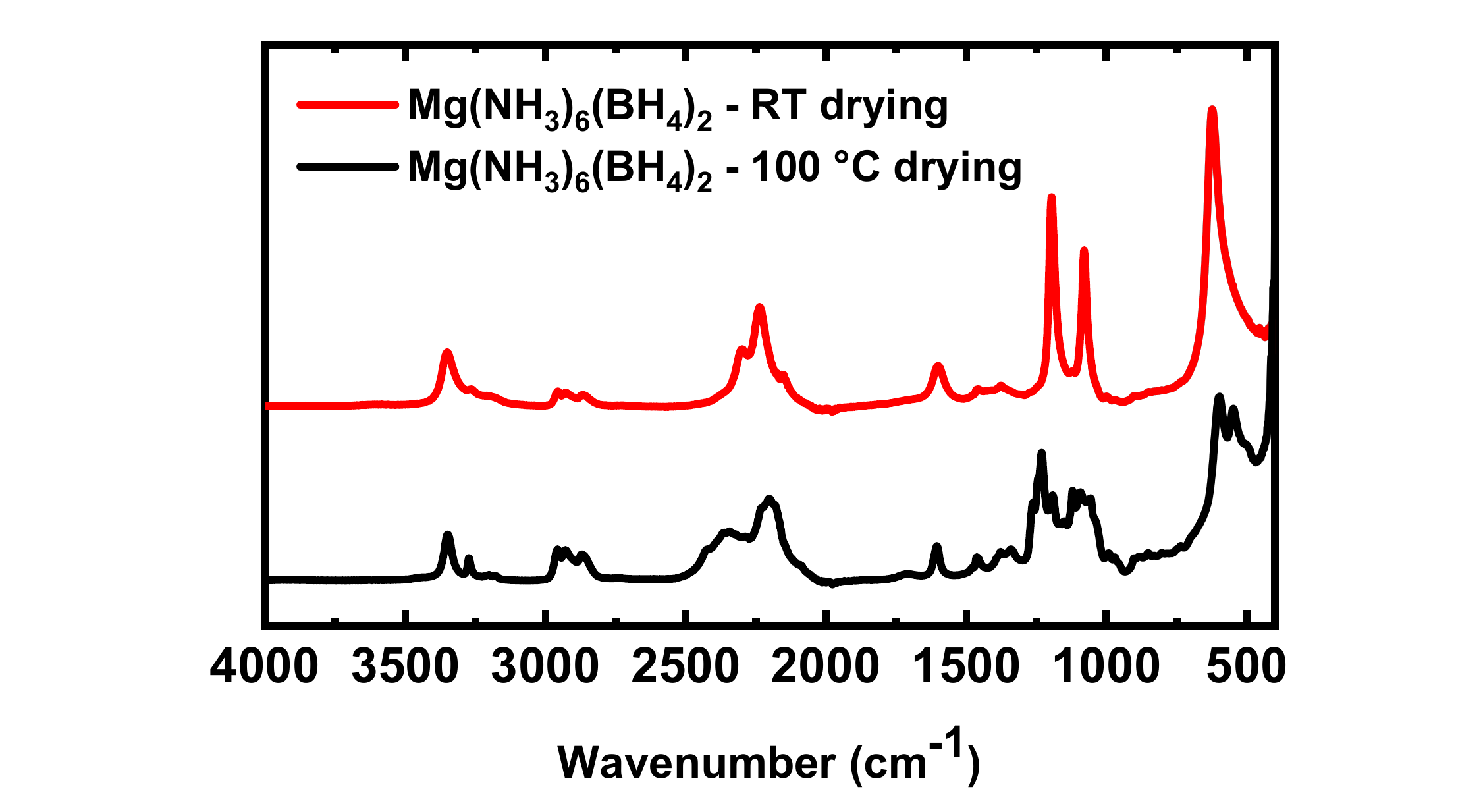}
    \caption{
        \label{Figure_S1}
        FTIR of Mg(NH$_3$)$_6$(BH$_4$)$_2$ showing the effect of drying post-synthesis at room-temperature vs 100°C. 
        Features due to the undesired diammoniate Mg(NH$_3$)$_2$(BH$_4$)$_2$ are more prominent during drying at 100°C.
    }
\end{figure}

\begin{figure}[h]
    \includegraphics[scale=0.96]{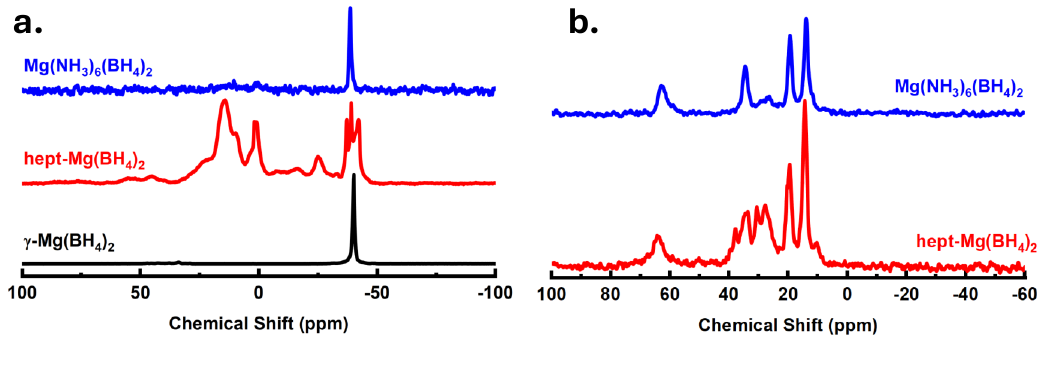}
    \caption{
        \label{Figure_S2}
        Cross-polarisation (a) $^{11}$B NMR on $\gamma$-Mg(BH$_4$)$_2$, hept-Mg(BH$_4$)$_2$, and Mg(NH$_3$)$_6$(BH$_4$)$_2$, and (b) $^{13}$C NMR data on hept-Mg(BH$_4$)$_2$ and Mg(NH$_3$)$_6$(BH$_4$)$_2$.
    }
\end{figure}

\begin{figure}[h]
    \includegraphics[scale=0.97]{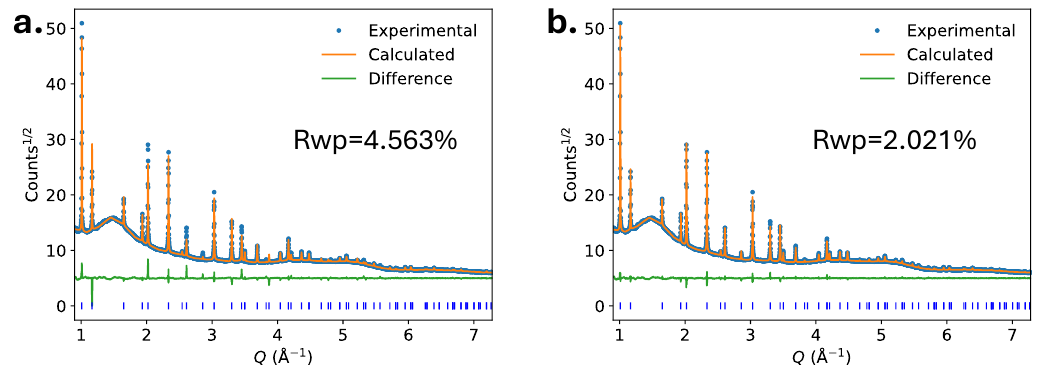}
    \caption{
        \label{Figure_S3}
        Rietveld refinements of the room-temperature BL2-1 diffraction data on Mg(NH$_3$)$_6$(BH$_4$)$_2$, using (a) a structure with the N anions on the high-symmetry sites by analogy with the Mg halide structure (i.e. without ammonia positional disorder) and (b) a structure with the N anions on low-symmetry, 25\% occupied sites as reported in Table~\ref{table1}.
    }
\end{figure}

\begin{figure}[h]
    \includegraphics[scale=1]{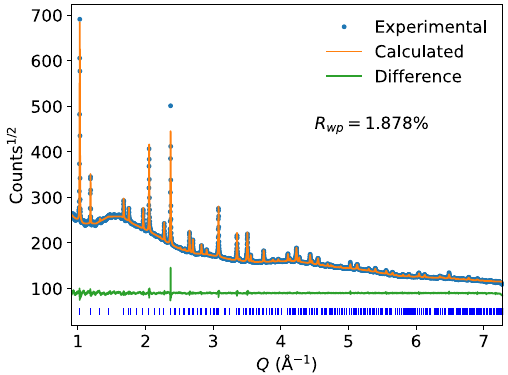}
    \caption{
        \label{Figure_S4}
        Pawley refinement of the 120\,K \textcolor{black}{beam line 2-1} diffraction data on Mg(NH$_3$)$_6$(BH$_4$)$_2$, using the \textcolor{black}{$Fm\bar{3}c$} space group.
    }
\end{figure}

\begin{figure}[h]
    \includegraphics[scale=0.95]{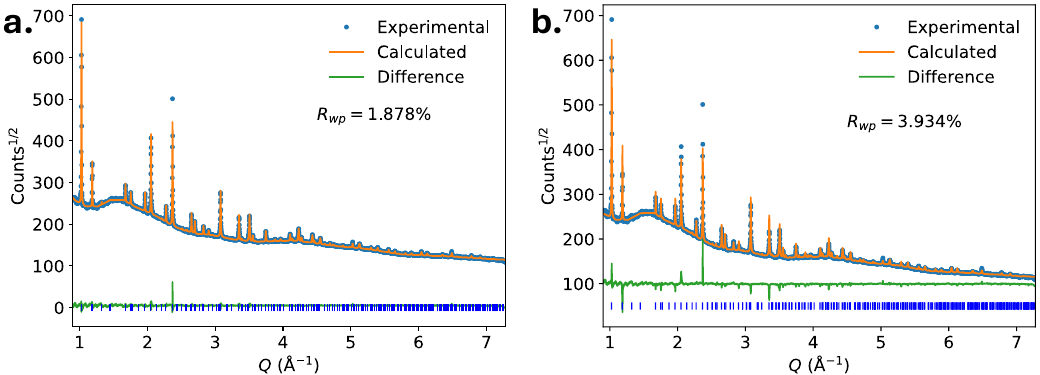}
    \caption{
        \label{Figure_S5.pdf}
        \textcolor{black}{
        (a) Pawley refinement and (b) Rietveld refinement of the 120\,K beam line 2-1 diffraction data on Mg(NH$_3$)$_6$(BH$_4$)$_2$, using the \textcolor{black}{$F\bar{4}3c$} space group, which loses an inversion center relative to the $Fm\bar{3}c$ space group (resulting in an additional degree of freedom, the N1 site $z$ coordinate). 
        This shows that removal of the inversion center does not enable a perfect fit. 
        }
    }
\end{figure}

\begin{figure}[h]
    \includegraphics[scale=1]{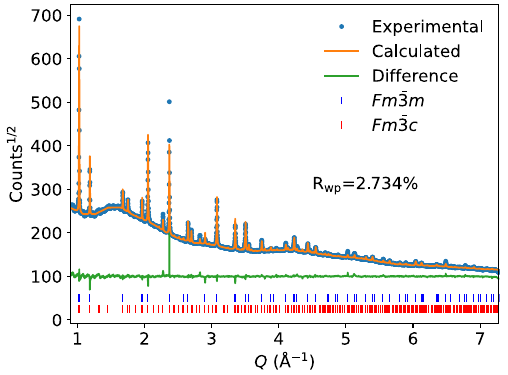}
    \caption{
        \label{SI_dual-Rietveld-120K.pdf}
        \textcolor{black}{
        A combined Rietveld refinement of the 120\,K beam line 2-1  diffraction data on Mg(NH$_3$)$_6$(BH$_4$)$_2$, assuming a mixture of the aristotype disordered $Fm\bar{3}m$ room-temperature structure and the tilt-ordered $Fm\bar{3}c$. 
        We can see that this dual-phase refinement does not succeed to fit the peakshape issue discussed in the main text. 
    }
    }
\end{figure}

\clearpage
\end{document}